\documentclass[prb,twocolumn,floatfix,superscriptaddress]{revtex4}
\usepackage{graphicx}
\usepackage{amsmath}

\begin{document}

\title{Molecular electronics and first-principles methods}
\author{J.~J.~Palacios}
\affiliation{Departamento de F\'{\i}sica Aplicada, Universidad de
Alicante, San Vicente del Raspeig, Alicante 03690, Spain.}
\author{A. J. P\'erez-Jim\'enez}
\affiliation{Departamento de Qu\'{\i}mica-F\'{\i}sica, Universidad de
Alicante, San Vicente del Raspeig, Alicante 03690, Spain.}
\author{E. Louis}
\affiliation{Departamento de F\'{\i}sica Aplicada, Universidad de
Alicante, San Vicente del Raspeig, Alicante 03690, Spain.}
\affiliation{Unidad Asociada del Consejo Superior de Investigaciones
Cient\'{\i}ficas, Universidad de
Alicante, San Vicente del Raspeig, Alicante 03690, Spain.}
\author{J. A. Verg\'es}
\affiliation{Departamento de Teor\'{\i}a de la Materia Condensada, Instituto
de Ciencia de Materiales de Madrid (CSIC), Cantoblanco, Madrid 28049, Spain.}
\affiliation{Unidad Asociada del Consejo Superior de Investigaciones
Cient\'{\i}ficas, Universidad de
Alicante, San Vicente del Raspeig, Alicante 03690, Spain.}
\author{E. SanFabi\'an}
\affiliation{Unidad Asociada del Consejo Superior de Investigaciones
Cient\'{\i}ficas, Universidad de
Alicante, San Vicente del Raspeig, Alicante 03690, Spain.}
\affiliation{Departamento de Qu\'{\i}mica-F\'{\i}sica, Universidad de
Alicante, San Vicente del Raspeig, Alicante 03690, Spain.}
\begin{abstract}
We discuss the key steps that have to be followed to calculate
coherent quantum transport in molecular and atomic-scale systems, making
emphasis on the {\it ab-initio}
Gaussian Embedded Cluster Method recently developed by the authors. We 
present various results on a simple system such as a clean
Au nanocontact and the same nanocontact in the presence of hydrogen
that illustrate the applicability of this 
method in the study and interpretation of
a large range of experiments in the field of molecular electronics.
\end{abstract}
\maketitle

\section{Introduction}
Nanoscale electronics constitutes the backbone of present
and future technological advances or, in other words, of Nanotechnology.
Ultimately, the functionality of electronic devices will rely on the conduction
properties of nanoscopic regions composed of a number of atoms that can range
typically from several thousands down to a {\em single} one. Probably, the
most promising research field within nanoscale electronics is what is known as
molecular electronics. The main idea behind molecular electronics 
is the possibility that functional units
can be built out of very stable and well-characterized
molecules\cite{Joachim:nature:01} such as fullerenes\cite{Dresselhaus:book:96}
or carbon nanotubes\cite{Saito:book:98,Dekker:pt:99}.
The simplest molecular device consists of two large
metallic electrodes, several nanometers apart, joined
by a molecule or molecules anchored to them\cite{Reed:science:97}.  We will
name this system molecular bridge.  Alternatively, the electrodes can be simply
connected by a chain of atoms of the same element as the
electrodes\cite{Muller:prl:92,Agrait:prb:93}.  These bridges, generically known
as atomic contacts or metallic nanocontacts, are not expected to be of any
practical  application, but constitute an excellent benchmark for us to learn
about the world of electrical transport at the atomic scale.

The design and fabrication of electronic devices at
the molecular and atomic scale poses new challenges for theorists who must
develop and implement new techniques to address the upcoming 
problems. The basics to calculate the zero-bias, zero-temperature
conductance $G$ of a molecular bridge or metallic nanocontact were
established by Landauer long before the concept of
molecular electronics was commonplace.  In Landauer's formalism $G$ 
is simply related to the quantum mechanical transmission probability 
$T$ of electrons at the Fermi level to go from one electrode to
the other\cite{Datta:book:95}(we assume spin degeneracy): 
\begin{equation}
G=G_0 T(E_{\rm F})= \frac{2e^2}{h} T(E_{\rm F}).
\label{g}
\end{equation}
The transmission can be easily estimated on generic
considerations for metallic nanocontacts\cite{Cuevas:prl:98:a,Hasmy:prl:01}, 
but it is much
more difficult to do so for molecular bridges.  The reason is simple: We do 
not know {\em a priori} 
where the electrode Fermi energy $E_{\rm F}$
lies with respect to the molecular levels. The
positioning of $E_{\rm F}$ depends on two closely-related factors:
(i) the coupling or hybridization
of the molecular levels with the free-electron levels in the 
electrodes and (ii) the charge transfer 
between molecule and electrodes. The conductance is thus strongly
dependent on the particular molecule, the detailed atomic arrangement of the
electrodes where the molecule binds, and the chemical nature of the various 
elements at play. In order
to give an answer to this problem we have to rely on first-principles
or {\em ab-initio} calculations.
 To calculate the atomic arrangement of the electrodes 
or the way the molecule binds to the electrodes is
a major problem in itself that deserves a whole study on its own and 
will not be given further consideration in these notes. 
Still, even if we ignore this important issue,
to implement Landauer formalism requires to know
the electronic structure of a finite region embedded in  
an infinite system with no particular 
symmetry or periodicity. This highly non-trivial problem is what we
discuss in what follows.

\section{The basics}
\begin{figure}
\includegraphics[width=3.0in,height=3.0in]{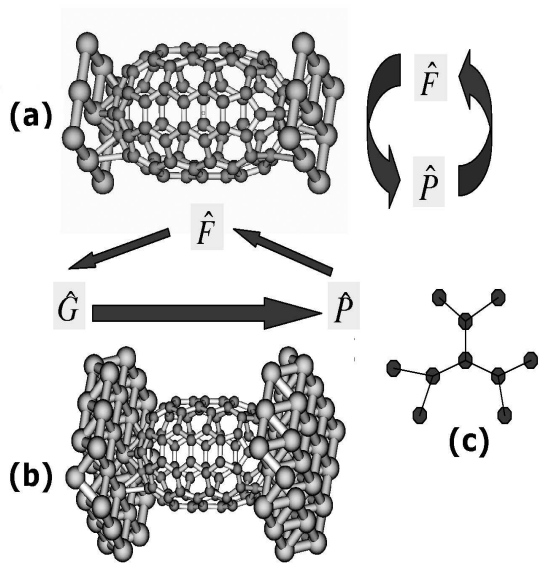}
\caption{(a) A  molecular bridge model where a 
short capped nanotube is contacted by two (001) parallel surfaces. 
The standard self-consistent procedure performed initially by 
GAUSSIAN98 is also shown schematically on the right. (b) The same system where
the first phantom atoms of the Bethe lattices are also depicted. 
The self-consistent procedure when the Bethe lattices are included
is also shown schematically on top. 
(c) A finite section of a Bethe lattice model
in two dimensions with coordination three. \label{cluster}}
\end{figure}

The main advantage of the numerical implementation that we discuss here 
with respect to similar proposals that have appeared in the 
literature\cite{Lang:prb:95,Hirose:prb:95,Brandbyge:prb:02,Taylor:prb:01:b}
is the use of a standard quantum
chemistry code such as GAUSSIAN98\cite{Gaussian:98} (see Refs.\ 
\onlinecite{Palacios:prb:01,Palacios:prb:02,Palacios:nano:01,Damle:prb:01}). 
The GAUSSIAN98 code provides a versatile method to
perfom first-principles or {\em ab-initio} 
calculations of clusters, incorporating the major
advancements in the field in terms of functionals, basis sets,
pseudopotentials, etc.. The procedure goes as follows.
A standard electronic structure calculation of the region that includes 
the molecule and a significant part of the electrodes is performed
[see Fig. \ref{cluster}(a)].  This calculation can be performed 
within any mean-field-like approach: Hartree-Fock or density functional 
(DF) theory in any of its multiple approximations. 
As far as transport is concerned,
the self-consistent hamiltonian $\hat H$ (or Fock matrix $\hat{F}$)
of this central cluster or supermolecule
contains the relevant information.
The retarded(advanced) Green's functions associated with  $\hat{F}$
\begin{equation}
\left [(E\pm i\delta)-\hat F\right ]\hat G^{(\pm)}= \hat I 
\end{equation}
needs to incorporate the rest of the infinite electrodes:
\begin{equation}
\left [(E\pm i\delta)-\hat F - \hat\Sigma^{(\pm)}
\right ] \hat G^{(\pm)}= \hat I 
\label{green}
\end{equation}
\noindent In this expression
\begin{equation}
\hat\Sigma^{(\pm)}=\hat\Sigma_{\rm R}^{(\pm)} +
 \hat\Sigma_{\rm L}^{(\pm)}, 
\end{equation}
\noindent where $\hat\Sigma_{\rm R}$($\hat\Sigma_{\rm L}$) denotes a 
self-energy operator
that accounts for the part of the right(left) semi-infinite electrode which has
not been included in the calculation,  and
$\hat I$ is the unity matrix. 
In a non-orthogonal basis the Green function takes the form
\begin{equation}
G_{NO}^{(\pm)}=S\left[(E\pm i\delta)S-F_{NO}-\Sigma^{(\pm)}_{NO}
\right ]^{-1}S
\end{equation}
where S is the overlap matrix. To simplify the notation we have dropped the
energy argument in the Green function and the selfenergy. 
The added self-energy matrices can only be
explicitly calculated in ideal situations. We 
choose to describe the bulk electrode with a Bethe lattice tight-binding model
with the coordination and parameters appropriate for  
the electrodes included in the cluster.
Figure \ref{cluster}(c) depicts schematically a Bethe lattice of coordination
three in two dimensions.  The
advantage of choosing a Bethe lattice resides in that it reproduces fairly well
the bulk density of states of most commonly used metallic electrodes.
For each atom in the outer planes of the
cluster, we choose to add a branch of the Bethe lattice in the direction of any
missing bulk atom (including those missing in the same plane).  In Fig.\
\ref{cluster}(b) the directions in which  Caylee tree
branches are added are indicated by 
bigger atoms which represent the first atom of the branch in that
direction. Assuming that the most important structural details of the electrode
are included in the central
cluster, the Bethe lattices should have no other relevance
than that of introducing a generic reservoir.

The above definition of the Green function is not a standard one. As
many other {\em ab-initio} implementations, GAUSSIAN98 makes use of 
non-orthogonal localized orbital basis sets.
The appearance of $S$ in the definition of the 
Green function guarantees that the total charge of the system $N$  can be
obtained through the standard expression used in quantum chemistry:
\begin{equation}
N={\rm Tr}[ P \cdot  S],
\end{equation}
where the trace runs over all the atomic orbitals of the cluster and
where $P$ is the density matrix defined in a standard way in terms 
of the retarded Green function as
\begin{equation}
P=-\frac{1}{\pi}\int_{-\infty}^{E_{\rm F}}{\rm Im}
\left[S^{-1} G^{(+)}_{\rm NO} S^{-1} \right ]
\label{eqn:nab}
\end{equation}
$E_{\rm F}$ is the Fermi energy which can be set by imposing overall
charge neutrality in the cluster.  
The integral in Eq. (\ref{eqn:nab}) can be efficiently
calculated along a contour in the complex 
plane\cite{Taylor:prb:01:b,Brandbyge:prb:02,Xue:condmat:01}. 
As explained in the next section, we force
GAUSSIAN98 to evaluate $F$ using the density matrix defined in 
Eq.~\ref{eqn:nab} 
instead of the standard one obtained from filling up the molecular orbitals
obtained from diagonalizing the Fock matrix\cite{Szabo:book:89}. 
It is also interesting to note that the
applicability of this approach, which we name the Gaussian Embedded
Cluster Method (GECM) can be easily implemented in any 
quantum-chemistry or {\em ab-initio}
code as long as it is based on a localized orbital expansion of the 
self-consistent, single-electron wavefunctions.

The transmission probability that appears in Eq. \ref{g} can be
calculated through the general expression\cite{Datta:book:95}
\begin{equation}
 T(E) =
{\rm Tr}[\hat\Gamma_{\rm L}\hat G^{(+)} \hat\Gamma_{\rm R}\hat G^{(-)}],
\end{equation}
where  Tr denotes the trace over all the orbitals of the cluster and 
where the operators $\hat\Gamma_{\rm R(L)}$
are given by $i(\hat\Sigma_{\rm R(L)}-\hat\Sigma^\dagger_{\rm R(L)})$. 
Finally, in order to single out the contribution of individual channels
to the current, one can diagonalize the transmission matrix. While the 
size of the matrix $[\hat\Gamma_L\hat G^{(-)} \hat\Gamma_R\hat G^{(+)}]$ can 
be as large as desired, the number of eigenvalues with a 
significant contribution will be typically much smaller, being determined by
the narrowest part of the nanocontact or the molecule.  The symmetry of
each channel can be identified by looking at its weight on
the atomic orbitals of the cluster.

\begin{figure}
\includegraphics[width=3.5in]{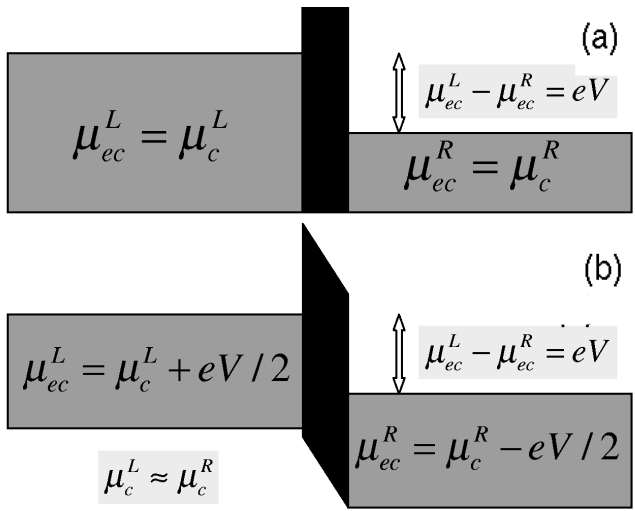}
\caption{Schematic picture of the development of a potential 
drop between electrodes separated by a barrier
on imposing an electrochemical potential difference between them. (a) Before
self-consistency begins the chemical potential in 
electrodes is different by an amount eV and the bottom of the conduction bands
is the same for both electrodes. This situation
corresponds to a diffusion problem. (b) 
After self-consistency has been achieved the chemical potential difference
practically vanishes in favor of an electrostatic potential difference between
eletrodes.
\label{drop}}
\end{figure}

Notice that, although the above expression provide us with
the transmission as a function of energy, only the value of $T$  
at the Fermi energy, $T(E_{\rm F})$,
is strictly correct (within the approximations inherent to DF theory).
Away from $E_{\rm F}$, $T(E)$ is only
indicative. For instance, if a gate voltage is applied and the 
system gets charged or
discharged, $T$ needs to be recalculated for the new Fermi energy.
To obtain the current $I$ at non-zero temperature and 
finite bias voltage $V$ one integrates the transmission
probability:
\begin{equation}
I=\frac{2e}{h}\int_{-\infty}^{\infty} T(E,V)
 [f_{\rm L}(E-eV/2)-f_{\rm R}(E+eV/2)] {\rm d}E,
\label{I}
\end{equation}
where $f_{\rm L}$ and $f_{\rm R}$ are
the Fermi distribution functions of the left and 
right electrodes, respectively. Notice, once again, that this innocent-looking
expression is not
a trivial generalization of Eq. \ref{g} extended to finite bias voltages and
temperature. The
transmission coefficient that appears in the integrand $T(E,V)$ depends on
temperature, but, most importantly,
{\em depends on $V$}. To calculate this transmission coefficient, knowledge of 
the Fock matrix in the presence of a bias voltage is required which, 
in turn, depends on the density matrix out of equilibrium. Based on
standard non-equilibrium Green function techniques, 
it has been repeatedly shown in the literature\cite{Datta:book:95} that
\begin{equation}
P=-\frac{i}{2\pi}\int_{-\infty}^{\infty}
{\rm d} E \left[S^{-1} G^{<} S^{-1} \right ]
\end{equation}
where 
\begin{equation}
\hat G^<(E)=i\hat G^{(-)}(E)\left[f_{\rm L}(E)\hat\Gamma_{\rm L}(E)+f_{\rm R}(E)\hat\Gamma_R(E)\right]\hat G^{(+)}(E).
\end{equation}
There are several technical and conceptual issues regarding the
computation of the above equations that need some discussion. We refer
the reader to the literature for a detailed account of most of 
them\cite{Taylor:prb:01:b,Palacios:prb:02,
Brandbyge:prb:02}. Here we will simply mention two of them. Maybe
the most important conceptual issue is that related to the fact that
using DF theory for non-equilibrium situations is not justified: DF theory
is a {\em ground-state} theory. 
We will make use of it simply because we do not know a better way to deal 
with out-of-equilibrium problems in an operative way.
Second, there is technical issue that, in our view,
has not received due attention in the
literature and that we find it worth a detailed 
discussion\cite{Louis:condmat:02}: The out-of-equilibrium electrostatics.
In most self-consistent 
approaches\cite{Lang:prb:95,Taylor:prb:01:b,
Damle:prb:01,Brandbyge:prb:02} one solves
the Poisson equation with boundary conditions appropriate for the electrode
geometries. This is equivalent to imposing an external electric field and
it can only be done for simple geometries such as infinite planes. 
The Keldysh formalism, however, does not necessarily require to deal with 
the Poisson equation
if a significant part of the  metallic electrodes has already
been included in the central cluster. Thus, one can consider realistic
electrode geometries if necessary. One simply imposes
an {\em electrochemical potential} difference between electrodes
 $\mu_{ec}^{\rm L}-\mu_{ec}^{\rm R}=eV$ which 
is actually what a battery does. There are two contributions to $\mu_{ec}$:
\begin{equation}
\mu_{ec}=\mu_{\rm chemical}+\mu_{\rm electrostatic},
\end{equation}
where we define
\begin{eqnarray}
\mu_{\rm chemical} &=& \lim_{\delta N \rightarrow 0} \delta E_{\rm core} / \delta N \nonumber \\
\mu_{\rm electrostatic}& =& \lim_{\delta N \rightarrow 0} \delta E_{ee} / \delta N. \nonumber
\end{eqnarray}
$E_{\rm core}$ is the standard energy contribution to the 
total energy from the cores of the atoms and $E_{ee}$ 
is the contribution from 
the electron-electron interaction\footnote{Note that this definition
is not a standard one in the sense that one usually thinks of external
fields when referring to the electrostatic potential contribution 
to the electrochemical potential. Instead we will make use of this term
when referring to any contribution coming from electron-electron interactions.
Strictly speaking, however,
one should not use the term electrostatic potential since
there are exchange and correlation contributions included in
DF theory.}.  Specifically
\begin{eqnarray}
\delta E_{\rm core} &=& {\rm tr}[\delta P\cdot h_{core}]\nonumber \\
\delta E_{\rm ee}& = &{\rm tr}[\delta P\cdot h_{ee}]\nonumber,
\end{eqnarray}
where $h_{core}$ and $h_{ee}$ are the core and interaction 
matrices composing the Fock
matrix.  In the selfconsistent procedure 
one electrode gets charged and the other discharged (by
the same amount in case of mirror symmetry). Once self-consistency 
has been achieved, an electrostatic potential difference $V$ between atoms one
or two layers inside opposite electrodes develops while their
chemical potentials are similar, i.e., they remain neutral. On the other hand,
the electrostatic potential drop will be
smaller than $V$ between atoms at the surface of opposite electrodes since they
carry the charges. We show this process schematically in  Fig. \ref{drop}.

\section{Example: Au nanocontacts}
\subsection{Clean nanocontacts}
\begin{figure}
\includegraphics[width=3.0in,height=2.5in]{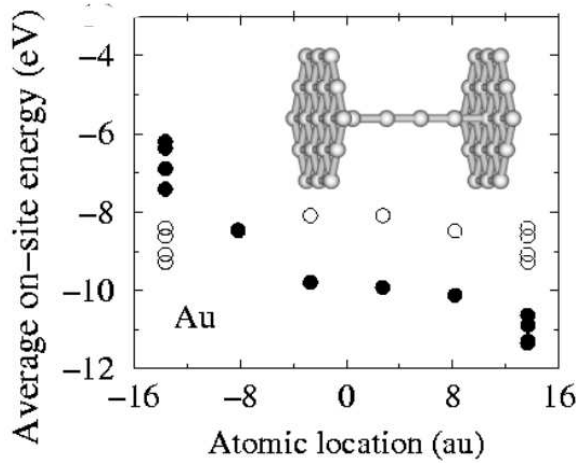}
\caption{Average on-site energies for all the atoms in the Au nanocontact 
shown in the inset. White dots have been obtained at zero bias and black dots
at 5 V.\label{Au-onsite}}
\end{figure}
\begin{figure}
\includegraphics[width=3.0in,height=2.5in]{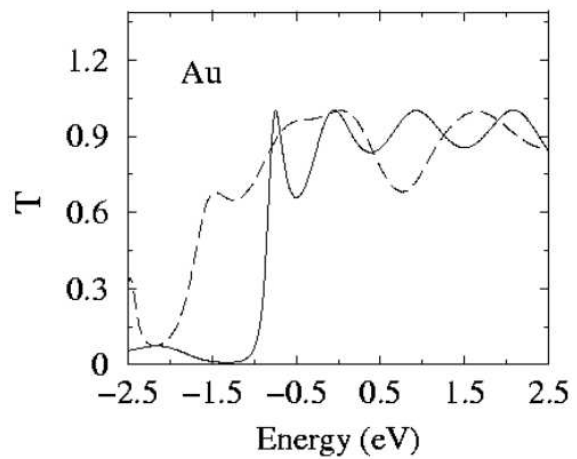}
\caption{Transmission versus energy for $V$=0 (continuous line)
and 5 V (dashed line) for the Au nanocontact shown in Figure 3. \label{Au-G}}
\end{figure}

Calculations that illustrate the methodology described above
have been carried out for the Au 
nanocontact shown in the inset of Fig.\ \ref{Au-onsite}:
Two (111) planes containing 19 atoms each plus a four-atom chain.
Interatomic bulk distances have been taken for the whole cluster (2.88\AA),
although one should keep in mind that an analysis of the structural
stability of the nanocontact model is an important issue when one is interested
in the interpretation of experimental data. The results shown next
correspond to zero temperature and 
for the DF calculations we have used the Becke's three-parameter 
hybrid functional using the Lee, Yang and Parr correlation functional 
(B3LYP)\cite{Becke:jcp:93} together with the semilocal shape consistent 
pseudopotential (SCPP) and minimal basis sets of 
Christiansen {\em et al.}\cite{Pacios:jcp:85,Hurley:jcp:86}.
Fig.\ \ref{Au-onsite} shows the average on-site energies of
the $5d 6s 6p$ orbitals for all the atoms of the nanocontact 
at zero and 5 V bias.
This magnitude reflects the local self-consistent
electrostatic potential on each atom.
The results represented by 
white dots correspond to zero bias where one can see the characteristic
potential profile due to the parallel surfaces
reflected on the chain atoms lying across the vacuum between
surfaces. The black dots correspond to a 5 V bias. One can see how 
an electrostatic potential drop
of almost 5 V has developed between plates {\em without having imposed
an external field, just an electrochemical potential difference}. 
 The major drop in the electrostatic 
potential occurs at the chain-plane contacts.
In the case of zero bias
the results are symmetric with respect to the geometric center of 
symmetry, as expected. Instead, a similar symmetry is absent for 5 V. 
Namely, whereas the potential drop between the left electrode
and the first atom in the chain is 1.92 eV, it is only of 0.85 eV between the 
chain end and the right electrode. 
This feature seems common to all previously reported 
results\cite{Brandbyge:prb:99,Brandbyge:prb:02}
and reflects bulk band structure features.
The transmission $T$ for zero and 5 V is shown in Fig.º\ref{Au-G}. 
$T$ oscillates around the Fermi energy with an upper limit value of one for
both cases. This limit is well documented theoretical and experimentally for
Au nanocontacts\cite{Agrait:pr:02}
since there is only one $s$-like channel around the Fermi ç
energy for Au chains. 
The oscillations are due to scattering at the electrode-chain contact. 
This explains why conductance histograms exhibit the lowest peak slightly
below $G=2e^2/h$\cite{Rego:condmat:02}.
It is worth noting  that although the total
current calculated by integrating $T(E,V=0)$ in a window [-$V$/2, $V$/2] 
does not differ much from that obtained with the full
non-equilibrium approach, the differential conductance
is significantly different. For instance, the gap below $-0.5$ eV is partially
filled at finite bias. 

\subsection{H$_2$ molecular bridges in Au nanocontacts}
\begin{figure}
\includegraphics[width=2.5in]{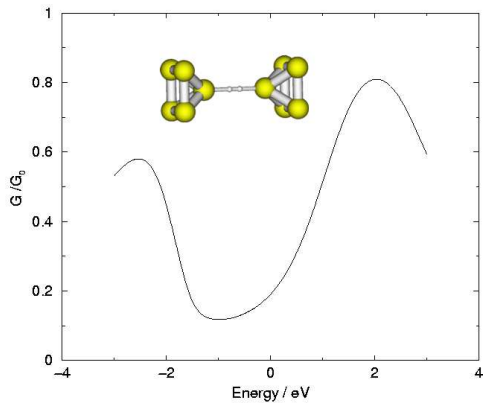}
\caption{Transmission versus energy for $V$=0 for the Au-H$_{2}$-Au bridge 
shown in the inset. \label{Au+H}}
\end{figure}

We finally examine the possible consequences of adding hydrogen (H$_2$)
to a Au nanocontact. A similar experimental
analysis has been recently reported for Pt with 
conclusive results: The presence of H$_2$ molecules alters
the conductance histograms of clean nanocontacts\cite{Smit:nature:02}. 
It is thus important 
to determine the extent of this influence when interpreting
conductance histograms under low-vacumm conditions or in the presence 
of gases. We considered here 
two electrodes in the form of pyramids in the (001) direction with a
H$_2$ molecule anchored between them (see inset in Fig.\ \ref{Au+H}). 
The orientation of the molecule has been chosen on the basis of the report
in Ref.\ \onlinecite{Smit:nature:02} for Pt, although other alternatives
have been recently proposed\footnote{Y. Garc\'{\i}a, J. J. Palacios, E. Louis,
J. A. Verg\'es, E. SanFabi\'an, and A. J. P\'erez-Jim\'enez, unpublished.}.
We have checked that this orientation is stable for a range of  distances $d$
between outer planes which are fixed in the relaxation.
Finally we have computed the conductance for $d=$8 \AA
(shown in Fig. \ref{Au+H}). As one can see the conductance at 
the Fermi energy is greatly reduced from the quantum of conductance $G_0$ due
to the presence of the H$_2$ molecule. A detailed analysis of this problem
will be published elsewhere\footnote{Y. Garcia, J. J. Palacios, 
A. J. Pº'erez-Jim\'enez, E. Louis, E. SanFabi\'an, and J. A. Verg\'es,
unpublished.}

Financial support by the spanish MCYT (grants BQU2001-0883,
PB96-0085, and MAT2002-04429-C03) and
the Universidad de Alicante is gratefully acknowledged.


\end{document}